# Quantum Structural Renormalization and Anharmonic Stabilization of Superconductivity in *P4/mmm* $YScH_8$

Xiaorui Dong[1†], Mi Pang[2†], Wenjie Ma[1*], Yao Ma[2*], Pugeng Hou[1*]

1 College of Science, Northeast Electric Power University, Changchun Road 169, 132012, Jilin, P. R. China
2 School of Optoelectronic Science and Intelligent Instrumentation & Shaanxi University Key Laboratory of Photonic Power Devices and Discharge Regulation, Xi'an University of Technology, Xi'an 710048, P. R. China

The recent synthesis of *P4/mmm*-$YScH_8$ at 191.7 GPa with a superconducting critical temperature $T_c$ of 113 K has attracted considerable interest in the study of ternary superhydrides. Here we systematically evaluate the effects of quantum and anharmonic motion of ions in *P4/mmm*-$YScH_8$ using the stochastic self-consistent harmonic approximation (SSCHA). We find that these effects renormalize the crystal structure and lower the dynamical stability threshold pressure from ~140 GPa to ~ 84 GPa, a reduction of about 40%. Furthermore, neglecting these effects causes a significant overestimation of the $T_c$, for instance by ~20 K at 190 GPa. Our prediction of $T_c$ at 190 GPa is 113 K ($\mu^* = 0.13$), close to the experimental observation. Analysis of the phonon density of states and Eliashberg spectral function further reveals site-dependent contributions to superconductivity. These results establish that a consistent treatment of nuclear quantum motion and anharmonicity is essential for accurately predicting superconductivity in compressed ternary hydrides.

[†] **Xiaorui Dong** and **Mi Pang** contribute equally to this work.

***Corresponding author:**
mawenjie@neepu.edu.cn
mayao@xaut.edu.cn
pugeng_hou@neepu.edu.cn

## I. INTRODUCTION

The quest for room-temperature superconductivity has motivated extensive research into hydrogen-rich materials under high pressure, following Ashcroft's seminal proposal that hydrogen-dominant materials could serve as high-temperature superconductors [1]. This strategy has led to remarkable successes, including the discoveries of superconductivity in $H_3S$ at 203 K [2] and $LaH_{10}$ at 260 K [3], which firmly established compressed hydrides as a viable platform for achieving high $T_c$. Building on these successes, more recently, attention has turned to ternary hydrides, which offer enhanced structural flexibility and the possibility of synergistic interstitial interactions that may further optimize superconducting properties [4, 5]. Among these, the Y–Sc–H system [6] has attracted particular interest due to the similar electronegativity and electron configurations of yttrium and scandium, which facilitate the formation of substitutional alloys.

Theoretical predictions by Shi et al. [6] using the harmonic approximation (on the classical structure) and first-principles calculations identified the *P4/mmm* phase of $YScH_8$ as a potential high-temperature superconductor, with a predicted $T_c$ of approximately 110 K at 140 GPa. In a very recent high-pressure experimental study, Jia et al. successfully synthesized this phase and reported a $T_c$ of 113 K at 191.7 GPa, along with structural stability down to at least 160 GPa [7]. A direct comparison reveals one striking discrepancy. In the vast majority of superconducting hydrides, increasing pressure hardens phonon modes and reduces the electronic density of states at the Fermi level, leading to a suppression of $T_c$. Even a nearly constant $T_c$ over such a wide pressure range, as observed in $H_3S$ [8], is considered highly unusual. A direct comparison between the harmonic prediction and the experimental measurement suggests that the conventional harmonic description fails to capture the pressure evolution of $T_c$. This near constancy (or slight increase) of $T_c$ over a 50 GPa pressure range strongly motivates the inclusion of anharmonic effects.

The limitations of the harmonic approximation for hydrogen-rich compounds under high pressure have been increasingly recognized in recent years. Pioneering work by Errea and colleagues demonstrated that quantum anharmonic effects can fundamentally reshape the phase diagrams of high-pressure hydrides [9-11]. In $H_3S$, quantum ionic fluctuations lower the hydrogen-bond symmetrization pressure by 72 GPa, stabilizing the high-symmetry *Im*-3*m* phase across the entire experimental window where the 203 K superconducting transition was observed [9, 10]. For $LaH_{10}$, quantum ionic fluctuations stabilize the highly symmetric *Fm*-3*m* structure down to 137 GPa — far below the harmonic instability limit of 230 GPa — yielding excellent agreement with the measured weakly pressure-dependent $T_c$ [11]. Building upon these seminal findings, anharmonicity has subsequently been shown to play a similarly critical role in other hydride systems, with effects that can either suppress or enhance $T_c$ depending on the specific bonding environment. For instance, in *Pm*-3*n*-$AlH_3$, anharmonicity hardens the low-energy optical modes, suppressing the electron-phonon coupling constant by more than 30% and strongly suppressing the superconductivity predicted by harmonic calculations, thereby resolving a long-standing disagreement between theory and experiment [12]. Similarly, in *Fm*-3*m* $NbH_3$, anharmonicity stabilizes the phase at pressures as low as 145 GPa — nearly 145 GPa lower than harmonic predictions — while also reversing the relative enthalpy ordering of competing phases [13]. For *Im*-3*m* $H_3Se$, anharmonicity stabilizes the structure down to approximately 61 GPa, much lower than the harmonic stability limit of 109 GPa, and significantly modifies the pressure dependence of $T_c$ [14]. Even in a $CH_4$-intercalated $H_3S$ system, anharmonic effects enable molecular rotation at room temperature, highlighting the profound impact of nuclear quantum effects on dynamical properties [15]. Collectively, these studies demonstrate that anharmonicity not only lowers the minimum pressure for dynamical stability — often by tens of GPa — but also fundamentally reshapes the pressure evolution of $T_c$. The specific manifestation of anharmonicity, whether it enhances or suppresses superconductivity, is highly system-dependent and cannot be reliably predicted by harmonic models alone. This necessitates a case-by-case, quantitative re-investigation of systems where harmonic predictions are at odds with experimental observations.

It is crucial to note that the harmonic predictions of Shi et al. were based on phonon calculations performed on the classically-optimized structure (i.e., obtained by minimizing the Born-Oppenheimer potential). However, upon explicitly including anharmonic effects via SSCHA, the equilibrium structure (hereafter "quantum structure", formally defined in Section IV) undergoes significant renormalization. The experimental $T_c$ of 113 K at 191.7 GPa is only marginally higher than the harmonic prediction of 110 K at 140 GPa, implying that the harmonic approximation may fails to capture the intrinsic pressure dependence of $T_c$ [6,7], we present a first-principles investigation that explicitly incorporates quantum ionic fluctuations and lattice anharmonicity via the stochastic self-consistent harmonic approximation (SSCHA). Our calculations show that even a seemingly minor difference between the quantum structure and the classical structure can significantly affect the $T_c$. And anharmonic effects substantially renormalize the phonon spectrum, particularly the hydrogen-dominated optical modes, and significantly alter the pressure range of dynamical stability. Moreover, we demonstrate that the anharmonic corrections naturally explain the pressure dependence of $T_c$, bringing theoretical predictions into much closer agreement with experimental observations [7]. These findings not only resolve the specific conflicts between theory and experiment for $YScH_8$ but also underscore the essential role of anharmonicity in accurately predicting superconductivity in ternary hydrides, thereby providing a refined theoretical framework for guiding future experimental and theoretical efforts in the search for high-temperature superconductors.

## II. METHODOLOGY

The influence of quantum ionic fluctuations and anharmonicity is assessed employing the stochastic self-consistent harmonic approximation (SSCHA) code [16], based on the earlier theoretical framework [17,18,19]. This section briefly outlines the SSCHA formalism and the computational approach used to compute the superconducting $T_c$, with anharmonic effects fully included.

### A. The stochastic self-consistent harmonic approximation

Without invoking any approximation to $V(\boldsymbol{R})$, the SSCHA accounts for quantum ionic fluctuations via a variational approach that minimizes the free energy, expressed with a trial density matrix $\tilde{\rho}_{\mathcal{R},\Phi}$ as

$$\mathcal{F}[\tilde{\rho}_{\mathcal{R},\Phi}] = < K + V(\boldsymbol{R}) > \tilde{\rho}_{\mathcal{R},\Phi} - TS\,[\tilde{\rho}_{\mathcal{R},\Phi}]. \quad (1)$$

Here, $K$ and $V(\boldsymbol{R})$ are the ionic kinetic energy and Born–Oppenheimer potential, respectively, while $T$ and $S[\tilde{\rho}_{\mathcal{R},\Phi}]$ are the temperature and entropy. The trial density matrix depends on the centroid positions $\boldsymbol{\mathcal{R}}$ (which fix the mean ionic coordinates) and auxiliary force constants $\boldsymbol{\Phi}$ (which govern the spread of the ionic wave functions about $\boldsymbol{\mathcal{R}}$). Minimizing $\mathcal{F}[\tilde{\rho}_{\mathcal{R},\Phi}]$ with respect to both $\boldsymbol{\mathcal{R}}$ and $\boldsymbol{\Phi}$ yields a robust variational free energy that requires no further approximations to $V(\boldsymbol{R})$. At the minimum, the equilibrium centroids $\boldsymbol{\mathcal{R}_{eq}}$ define the average ionic positions, while $\boldsymbol{\Phi_{eq}}$ characterizes the associated quantum fluctuations. In the classical harmonic picture, by contrast, the equilibrium geometry $\boldsymbol{R}_0$ is obtained by directly minimizing $V(\boldsymbol{R})$; these positions generally deviate from $\boldsymbol{\mathcal{R}_{eq}}$ because the ionic kinetic energy is entirely neglected. The SSCHA code also handles structural optimization—including lattice degrees of freedom—while fully including quantum and anharmonic effects at any prescribed pressure.

In the static limit [16], phonon frequencies are obtained from the eigenvalues of the mass-rescaled second-order derivatives of the free energy with respect to the centroid positions, evaluated at $\boldsymbol{\mathcal{R}_{eq}}$.

$$\boldsymbol{D}_{\boldsymbol{ab}}^{(F)} = \frac{1}{\sqrt{M_a M_b}} \frac{\partial^2 F}{\partial \boldsymbol{R}_a \partial \boldsymbol{R}_b} |_{\boldsymbol{R}_{eq}}. \quad (2)$$

Where $a$ and $b$ are composite indices spanning atomic sites and Cartesian directions, and $M_a$ is the mass of atom $a$. This dynamical matrix $\boldsymbol{D}^{(F)}$—often referred to as the free energy Hessian–is the quantum anharmonic analogue of the conventional harmonic dynamical matrix, which is instead constructed from the Hessian of $V(\boldsymbol{R})$ at $\boldsymbol{R}_0$:

$$\boldsymbol{D}_{\boldsymbol{ab}}^{(h)} = \frac{1}{\sqrt{M_a M_b}} \frac{\partial^2 V(\boldsymbol{R})}{\partial R_a \partial R_b} |_{\boldsymbol{R}_0}. \quad (3)$$

Notably, negative eigenvalues of $\boldsymbol{D}^{(F)}$ signal a structural instability within the anharmonic free-energy landscape, whereas negative eigenvalues of $\boldsymbol{D}^{(h)}$ indicate an instability when quantum ionic effects are disregarded.

Beyond optimizing internal atomic positions via minimization of $\mathcal{F}[\tilde{\rho}_{\mathcal{R},\Phi}]$ with respect to the centroids, SSCHA also relaxes the lattice parameters under anharmonic and quantum effects. This is done by computing the stress tensor via the derivative of the SSCHA free energy with respect to the strain tensor components $\epsilon$:

$$P_{\alpha\beta} = -\frac{1}{\Omega_V} \frac{\partial \mathcal{F}}{\partial_{\epsilon_{\alpha\beta}}} |_{\epsilon=0}. \quad (4)$$

Where $\Omega_V$ is the simulation cell volume [16], and $\alpha$, $\beta$ are Cartesian components. This formulation naturally includes the additional pressure arising from ionic fluctuations, beyond the classical harmonic pressure obtained by replacing $\mathcal{F}$ with $V(\boldsymbol{R})$ in Eq. (4). We adopt the static limit of the SSCHA dynamical theory, which has been shown to provide accurate phonon frequencies and EPC parameters for similar strongly anharmonic hydrides [10-19].

**B. Calculation of the superconducting transition temperature**

We evaluated the $T_c$ with the Allen-Dynes modified McMillan equation [20],

$$T_c = \frac{f_1 f_2 \omega_{log}}{1.2} \exp\left[-\frac{1.04(1+\lambda)}{\lambda-\mu^*(1+0.62\lambda)}\right]. \tag{5}$$

Where $\lambda$ represents the electron-phonon coupling constant, and $\mu^*$ is the Coulomb pseudopotential [20]. In the Allen-Dynes formula, the Coulomb pseudopotential $\mu^*$ is a semi-empirical parameter that accounts for the repulsive electron-electron interaction. For conventional superconductors, $\mu^*$ typically ranges from 0.10 to 0.13. In the study of high-pressure hydrides, a value of $\mu^* = 0.13$ is commonly adopted, as it has been shown to yield $T_c$ values in good agreement with experimental measurements for a wide range of superconducting hydrides. Therefore, we present our main results using $\mu^* = 0.13$, and also provide calculations with $\mu^* = 0.10$ for comparison. These equations have yielded $T_c$ values that agree well with experimental results in super-hydrides despite its simplicity [10, 11]. $\lambda$ is defined as the first reciprocal moment of the electron-phonon Eliashberg function $\alpha^2F(\omega)$,

$$\lambda = 2\int_0^\infty \frac{\alpha^2F(\omega)}{\omega} d\omega. \tag{6}$$

The other parameters in Eqs. (5) are calculated as follows:

$$\omega_{log} = \exp\left[\frac{2}{\lambda}\int_0^\infty \frac{d\omega}{\omega}\alpha^2F(\omega)\ln\omega\right], \tag{7}$$

$$f_1 = \sqrt[3]{\left[1+\left(\frac{\lambda}{\Lambda_1}\right)^{\frac{3}{2}}\right]}, \tag{8}$$

$$f_2 = 1 + \frac{\left(\frac{\bar{\omega}_2}{\omega_{log}}-1\right)\lambda^2}{\lambda^2+{\Lambda_2}^2}. \tag{9}$$

$\Lambda_1$, $\Lambda_2$ and $\bar{\omega}_2$ are given by

$$\Lambda_1 = 2.46(1+3.8\mu^*), \tag{10}$$

$$\Lambda_2 = 1.82(1+6.3\mu^*)\frac{\bar{\omega}_2}{\omega_{log}}, \tag{11}$$

$$\bar{\omega}_2 = \sqrt{\frac{2}{\lambda}\int_0^\infty \alpha^2F(\omega)\omega\, d\omega}. \tag{12}$$

We calculate the Eliashberg function as

$$\alpha^2F(\omega) = \frac{1}{2N(0)N_qN_k}\sum_{knm,\mu q,\bar{a}\bar{b}} \frac{\epsilon_\mu^{\bar{a}}(\boldsymbol{q})\epsilon_\mu^{\bar{b}}(\boldsymbol{q})^*}{\omega_\mu(\boldsymbol{q})\sqrt{M_{\bar{a}}M_{\bar{b}}}} \times d^{\bar{a}}_{\boldsymbol{k}n,\boldsymbol{k}+\boldsymbol{q}m} d^{\bar{b}*}_{\boldsymbol{k}n,\boldsymbol{k}+\boldsymbol{q}m}\delta(\varepsilon_{kn})\delta(\varepsilon_{k+qm})\delta(\omega-\omega_\mu(\boldsymbol{q})). \tag{13}$$

In the equation above, $d^{\bar{a}}_{\boldsymbol{k}n,\boldsymbol{k}+\boldsymbol{q},m} = \left< \boldsymbol{k}n \left| \frac{\delta V_{KS}}{\delta R^{\bar{a}}(\boldsymbol{q})} \right| \boldsymbol{k}+\boldsymbol{q}, m \right>$, where $|\boldsymbol{k}n>$ represents a Kohn-Sham state with energy $\varepsilon_{\boldsymbol{k}n}$ measured from the Fermi level, $V_{KS}$ denotes the Kohn-Sham potential, and $R^{\bar{a}}(\boldsymbol{q})$ stands for the Fourier-transformed displacement of atom $\bar{a}$; The combined atom and Cartesian indices with a bar ($\bar{a}$) only run for atoms within the unit cell. $N_k$ and $N_q$ are the number of electron and phonon momentum points utilized for BZ sampling; $N(0)$ represents the density of states at the Fermi level, while $\omega_\mu(\boldsymbol{q})$ and $\epsilon_\mu^{\bar{a}}(\boldsymbol{q})$ represent phonon frequencies and the polarization

vectors, respectively. In this study, the Eliashberg function is computed both at the harmonic and anharmonic levels, by substituting into Eqs. (13). The harmonic phonon frequencies and polarization vectors obtained by diagonalizing $\boldsymbol{D}^{(h)}$ or their anharmonic equivalents from diagonalizing $\boldsymbol{D}^{(F)}$. It is important to note that the derivatives of the Kohn–Sham potential used in the electron–phonon matrix elements are evaluated at different positions in classical harmonic and quantum anharmonic calculations: in the former, they are computed at the positions $\boldsymbol{R}_0$ minimizing $V(\boldsymbol{R})$, while in the latter, they are determined at the positions $\boldsymbol{\mathcal{R}}_{eq}$ that minimize instead $\mathcal{F}[\tilde{\rho}_{\mathcal{R},\Phi}]$. For comparison, $T_c$ is also determined from solving the isotropic Migdal-Eliashberg equations once $\alpha^2F(\omega)$ is obtained [21].

# III. COMPUTATIONAL DETAILS

The ab initio calculations were performed using the QUANTUM ESPRESSO (QE) package [22], employing ultrasoft pseudopotentials [23] with the Perdew-Burke-Ernzerhof (PBE) parametrization [24] of the exchange correlation potential. The plane-wave basis cutoff was set to 80 Ry and 800 Ry for the density. For the electronic Brillouin-zone integrations, a 20 × 20 × 12 Monkhorst-Pack grid [25] with a smearing parameter of 0.02 Ry was employed. A 6 × 6 × 3 q-point mesh was used for the phonon (or electron–phonon) calculations. SSCHA minimization [16] requires the calculation of energies, forces, and stress tensors in supercells. These calculations were conducted within DFT [26] at the PBE level using Quantum ESPRESSO, employing the same pseudopotentials. We performed the calculations in a 2 × 2 × 1 supercell containing 40 atoms, resulting in dynamical matrices on a commensurate 2 × 2 × 1 grid. A 60 Ry energy cutoff and a 6 × 6 × 4 k-point mesh for Brillouin zone (BZ) integrations were sufficient to converge the SSCHA gradient in the supercell. The SSCHA calculations were performed at 0 K. Following each minimization iteration, we augmented the population with a greater number of individuals N using the minimized trial density matrix until convergence was achieved. Two criteria were employed to terminate the minimization loops: firstly, a Kong-Liu ratio, which assesses the effective sample size and should attain a value of 0.5, and secondly, a ratio of less than $10^{-9}$ between the free energy gradient and its stochastic error. The difference between the harmonic and anharmonic dynamical matrices on the 2 × 2 × 1 grid was interpolated to a 6 × 6 × 3 grid. By summing the harmonic dynamical matrices in this fine grid to the result, the anharmonic dynamical matrices on the 6 × 6 × 3 grid were obtained. For electronic integration in Eq. (13), a 30 × 30 × 30 k-point grid was employed, and the Dirac deltas were approximated with Gaussian functions with a width of 0.012 Ry. Convergence tests with respect to the k-point mesh (up to 40×40×40) and Gaussian broadening (from 0.004 to 0.100 Ry) were carefully performed.

## IV. RESULTS AND DISCUSSION

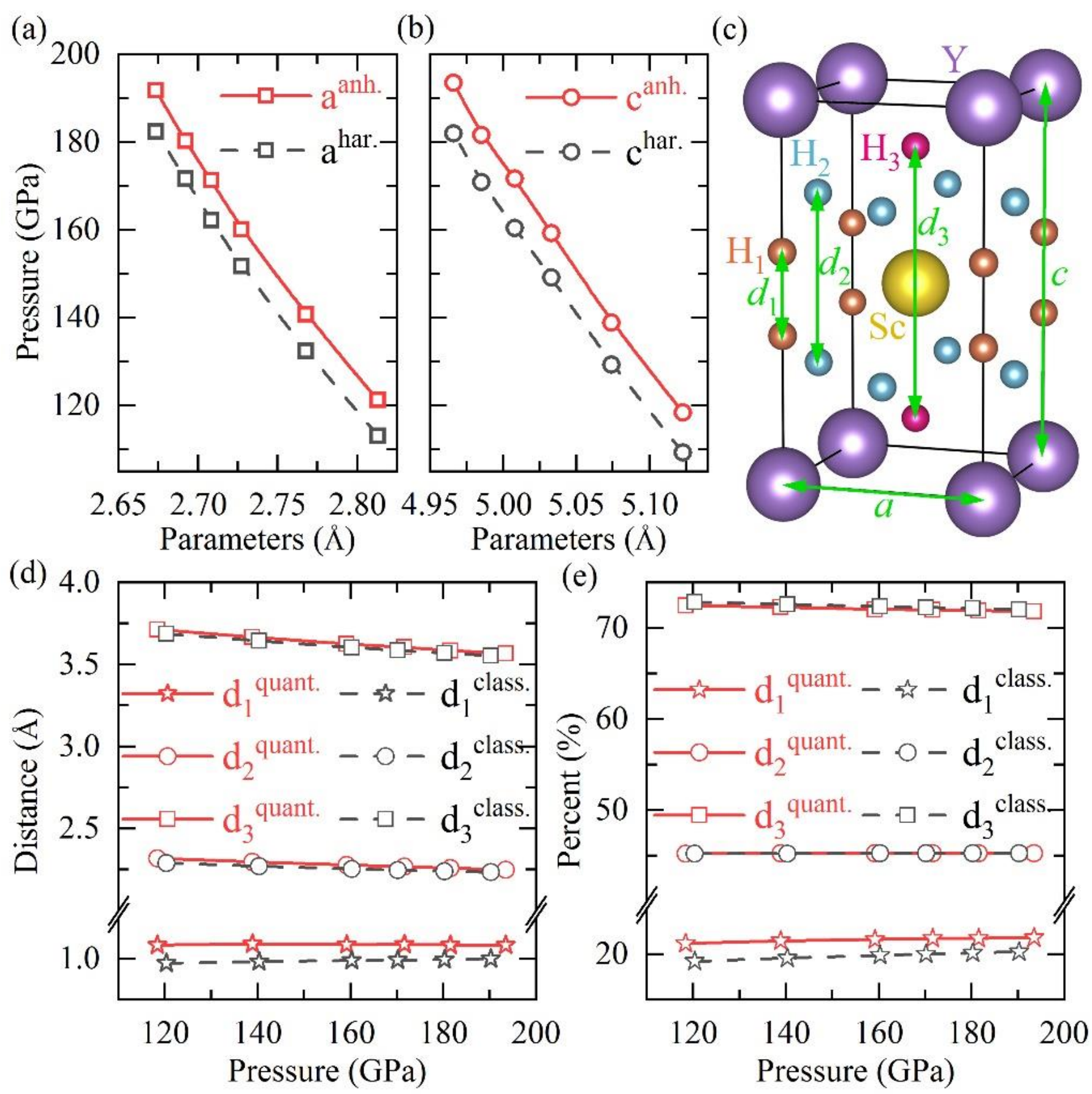

FIG. 1. (a) – (b) Comparison of harmonic and anharmonic pressures as functions of lattice parameters *a* and *c*, both evaluated on the quantum structure. (c) The crystal structure of *P*4/*mmm*-$YScH_8$ is shown with the unit cell, where middle golden and big purple spheres represent Sc and Y atoms, respectively. The three hydrogen sites are labeled as H1 (small orange, corresponding to $d_1$), H2 (small blue, corresponding to $d_2$), and H3 (small magenta, corresponding to $d_3$). Along the c-axis direction, the H–H distances for the pairs located at the edge, on the face, and inside the unit cell are denoted as $d_1$, $d_2$ and $d_3$, respectively. (d) – (e) Comparison of structural parameters between the classical structure (harmonic) and the quantum structure (anharmonic) as functions of pressure. In (d) and (e), $d_1$, $d_2$, $d_3$ are distinguished by markers (stars, circles, squares) as shown in the legend.

Within the classical harmonic approximation, the crystal structure is determined by minimizing the Born–Oppenheimer potential energy surface $V(\boldsymbol{R})$. We refer to this structure as the "classical structure," which serves as the basis for the harmonic calculations in previous studies [6]. When quantum ionic fluctuations and lattice anharmonicity are explicitly included via the stochastic self-consistent harmonic approximation (SSCHA) [16], the equilibrium structure is instead obtained by minimizing the free energy $\mathcal{F}[\tilde{\rho}_{\mathcal{R},\Phi}]$ at 0 K, which incorporates the vibrational contribution. This anharmonically renormalized structure is termed the "quantum structure." Throughout this work, we distinguish among three types of calculations:

(A) Harmonic on classical: Harmonic phonons computed on the classical structure (reproducing Shi et al. [6]).

(B) Harmonic on quantum: Harmonic phonons computed on the quantum structure (our own harmonic reference).
(C) Anharmonic on quantum: Anharmonic phonons from the SSCHA free-energy Hessian on the quantum structure (our main result).
All harmonic and anharmonic results presented in the following sections, unless otherwise specified, are obtained based on this quantum structure using either method (B) or (C). For completeness, we have also verified that our harmonic calculations on the classical structure (method A) reproduce the results of Shi et al. exactly (e.g., $T_c$ = 109.8 K at 140 GPa), as discussed later. Along the $c$-axis direction, three characteristic H–H distances are denoted as $d_1$ (edge), $d_2$ (face), and $d_3$ (inside the unit cell).

Fig. 1(a) and 1(b) compare the classical pressure (from the harmonic approximation) and the quantum pressure (from SSCHA) as functions of the lattice parameters $a$ and $c$, respectively. Both pressures are evaluated on the same quantum structure, highlighting the intrinsic pressure renormalization arising from quantum ionic fluctuations at a fixed geometry, independent of any structural change. For the same set of lattice parameters ($a$ = 2.673 Å, $c$ = 4.966 Å), the quantum pressure is systematically higher than the classical one, with a difference of approximately 8 GPa. This quantum correction to the equation of state is a general feature of compressed superhydrides[11,12,13], arising from quantum and anharmonic effect of hydrogen atoms. Consequently, to compare with experimental pressures (which are inherently quantum), the harmonic approximation would require a different lattice parameter from the SSCHA method for the same nominal pressure.

Figures 1(d) and 1(e) present the evolution of the three H–H distances $d_1$, $d_2$, $d_3$ and their ratios to the $c$-axis lattice parameter as a function of pressure, comparing the classical structure (harmonic optimization) with the quantum structure (SSCHA optimization). The most pronounced structural change between the classical and quantum structures occurs for $d_1$, (the H–H pair at the edge of the unit cell). This significant renormalization of the local geometry, particularly for the edge H-H pair, is a direct consequence of the anharmonic free-energy minimization and forms the microscopic origin for the modified phonon spectra discussed below. At 140 GPa, for example, anharmonicity increases $d_1$ from approximately 0.981 Å (harmonic) to 1.089 Å (anharmonic), an increase of about 0.108 Å, whereas the changes in $d_2$ and $d_3$ are only about 0.02 Å. Consequently, the anharmonic correction to $d_1$ is more than five times larger than those to $d_2$ or $d_3$. When expressed as fractions of the $c$-axis parameter [Fig. 1(e)], anharmonic effects raise the ratio $d_1/c$ from about 19.5% (harmonic) to 21.5% (anharmonic) at 140 GPa – an increase of 2% – while the ratios $d_2/c$ and $d_3/c$ exhibit negligible variations (~0.2%). At all pressures studied, anharmonic effects impose a similar structural distortion, and the magnitude of these corrections tends to decrease with increasing pressure, consistent with the behavior observed in other high-pressure hydrides such as $CH_4$-$H_3S$[15], $ScH_6$[27] and $SnH_4$[28].

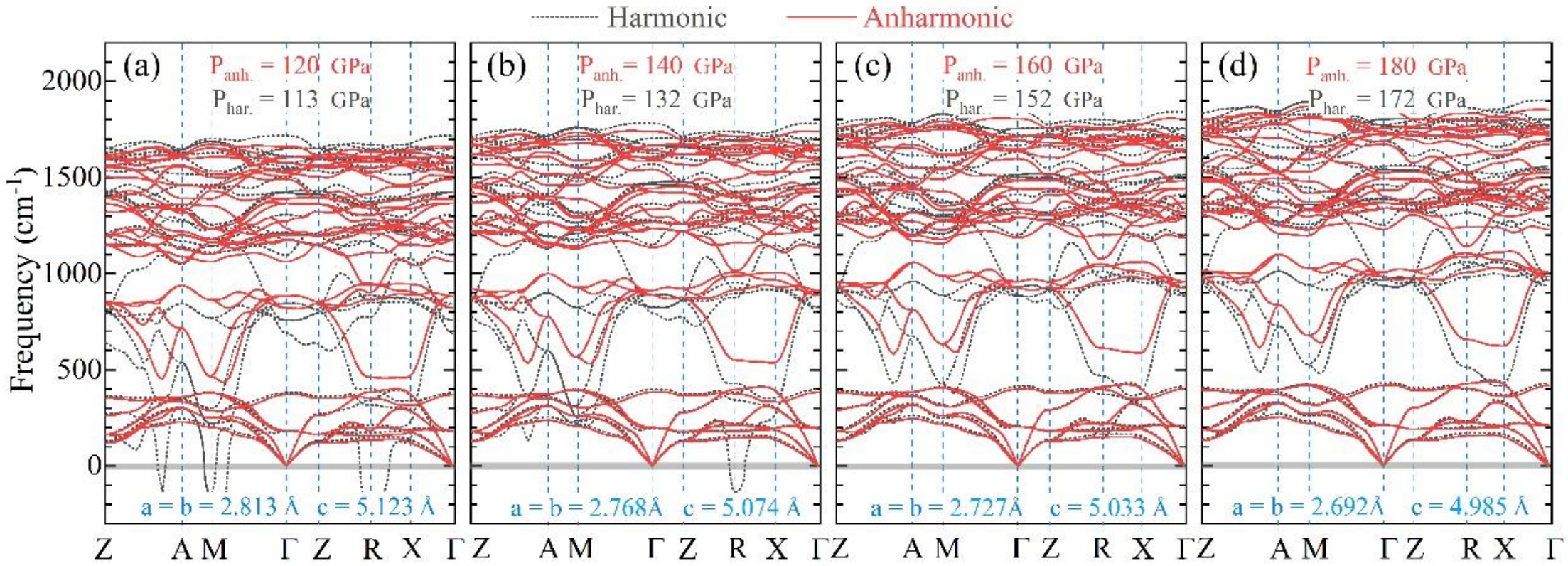

FIG. 2. Comparison between the harmonic (evaluated on the anharmonically optimized "quantum structure", black dotted lines) and anharmonic (from the free-energy Hessian, red solid lines) phonon spectra at (a, c) = (2.813, 5.123), (2.768, 5.074), (2.727, 5.033), (2.692, 4.985) Å for panels (a)–(d), respectively, along the high-symmetry path Z–A–M–Γ–Z–R–X–Γ. The anharmonic spectra are obtained from $\boldsymbol{D}^{(F)}$ and correspond to the static limit of the SSCHA dynamical theory. The pressures from the harmonic calculation on the quantum structure and from the anharmonic calculation are marked at the top of each panel: (a) $P_{anh.}$ = 120 GPa, $P_{har.}$ = 113 GPa; (b) $P_{anh.}$ = 140 GPa, $P_{har.}$ = 132 GPa; (c) $P_{anh.}$ = 160 GPa, $P_{har.}$ = 152 GPa; (d) $P_{anh.}$ = 180 GPa, $P_{har.}$ = 172 GPa. The harmonic and anharmonic pressures differ because the same lattice parameters yield different effective pressures in the two frameworks ($P_{anh} \approx P_{har} + 8$ GPa). The grey solid line at zero frequency separates positive (real) frequencies from negative (imaginary) frequencies.

Figure 2 presents a comparison between the harmonic (method B, black dotted lines) and anharmonic (method C, red solid lines) phonon spectra of the *P*4/*mmm* phase of $YScH_8$ along the high-symmetry path Z–A–M–Γ–Z–R–X–Γ, calculated at four different sets of lattice parameters corresponding to panels (a)–(d). The anharmonic spectra are obtained from the free-energy Hessian $\boldsymbol{D}^{(F)}$ within the static limit of the SSCHA dynamical theory, while the harmonic ones derive from the conventional second-order expansion of the Born-Oppenheimer potential. The top of each panel reports the harmonic pressure (from method B) and the anharmonic pressure (from method C) evaluated for the same quantum crystal structure: (a) 113 vs. 120 GPa, (b) 132 vs. 140 GPa, (c) 152 vs. 160 GPa, and (d) 172 vs. 180 GPa. A horizontal grey solid line at zero frequency separates the region of positive (real) frequencies from that of negative (imaginary) ones, thereby providing a direct visual criterion for dynamical stability. It should be noted that the present SSCHA calculations address dynamical stability only. The thermodynamic stability of *P*4/*mmm* relative to possible decomposition into $YH_4$, $ScH_6$, or elemental phases in this pressure range warrants future investigation.

Several important features emerge from the comparison. Within the harmonic approximation on the quantum structure (method B), the system exhibits pronounced imaginary frequencies in panels (a) and (b), indicating dynamical instability at pressures below approximately 132GPa. In panels (c) and (d), the harmonic spectra become fully positive, implying that the harmonic stability threshold on the quantum structure is no higher than 152 GPa. To further confirm the harmonic stability pressure, we performed a quadratic fit of the squared frequencies, obtaining a pressure of 145 GPa at which the lowest frequency (or its square) at the high-symmetry *R* point in the Brillouin zone becomes exactly zero. This value (method B) is, however, slightly higher than the previously reported harmonic threshold of ~140 GPa (method A), which was obtained using the classical structure [6]. For completeness, we have verified that our harmonic phonon calculations performed on the classical structure reproduce the previous results exactly, with the instability appearing below ~120 GPa. The stark difference in the harmonic stability threshold between the classical and quantum structures underscores the profound impact of anharmonicity on the structural optimization itself. In contrast, the anharmonic spectra obtained from the SSCHA remain completely devoid of imaginary frequencies in all four panels, from 120 GPa up to 180 GPa. More importantly, further anharmonic calculations carried out at even lower pressures (not shown in the figure) reveal that the *P*4*/mmm* structure remains dynamically stable down to as low as ~84 GPa, as determined by a quadratic fit of the lowest squared phonon frequency versus pressure. It is important to stress, however, that this ~84 GPa boundary refers to the dynamical stability limit of the *P*4*/mmm* lattice, rather than the thermodynamic one. The experimental observation of the phase down to only ~160 GPa does not necessarily imply a breakdown of the structure at lower pressures; instead, the absence of synthesis below this threshold could plausibly be associated with kinetic barriers, competing phases, or intrinsic constraints of the high-pressure synthesis route. This substantial lowering of the stability boundary by more than 60 GPa relative to the harmonic prediction highlights the decisive role of quantum ionic fluctuations and lattice anharmonicity. The present results bring the theoretical stability boundary into much closer agreement with the experimental observation of stability down to ~160 GPa, in contrast to the harmonic prediction that places the stability limit at a much higher pressure, and further demonstrate that the anharmonic stabilization extends to a pressure regime that has not yet been explored experimentally [7], providing a solid foundation for the subsequent analysis of the superconducting properties.

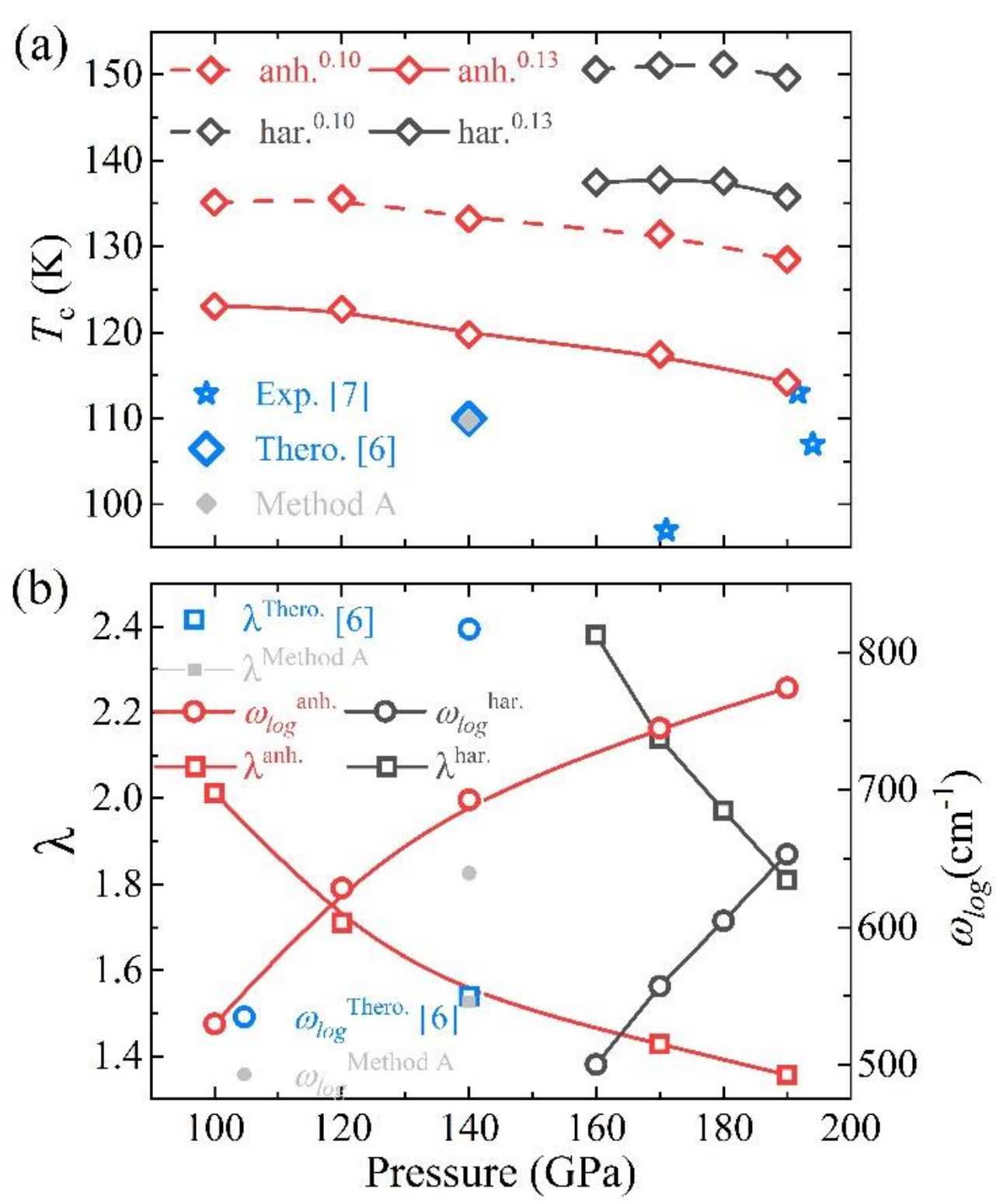

FIG. 3. Superconducting properties calculated on the anharmonically optimized "quantum structure" and on the “classical structure” within the harmonic approximation. (a) Comparison of the superconducting critical temperature

$T_c$ between harmonic (black lines) and anharmonic (red lines) evaluated using the Allen-Dynes modified McMillan equation as a function of pressure. $T_c$ values calculated with $\mu^*$ = 0.13 (solid lines) and $\mu^*$ = 0.10 (dashed lines) are shown; diamonds mark the calculated data points. The harmonic results here are obtained from the quantum structure, which differ from the classical structure harmonic results of Shi et al. [6] (blue diamonds). The experimental data from Jia et al. [7] are shown as blue stars. (b) Comparison of the electron-phonon coupling constant λ (square symbols) and the average logarithmic frequency $\omega_{log}$ (circle symbols) between harmonic (black lines) and anharmonic (red lines) as a function of pressure. For both harmonic and anharmonic results (both evaluated on the quantum structure), lines connect the data points. The results from Shi et al. [6] (on the classical structure) are shown as blue squares (λ) and blue circles ($\omega_{log}$). Our harmonic calculations based on the classical structure are represented by gray symbols in (a) and (b): diamonds for $T_c$, squares for λ, and circles for $\omega_{log}$.

Figure 3(a) compares the pressure dependence of the superconducting critical temperature $T_c$ for the *P4/mmm* phase of $YScH_8$, evaluated within both the harmonic and anharmonic frameworks using the Allen‑Dynes modified McMillan equation. The anharmonic results (red lines) are obtained from the free-energy Hessian phonons within the static limit of the SSCHA, while the harmonic ones (black lines) are derived from the conventional second-order expansion of the Born‑Oppenheimer potential. With a Coulomb pseudopotential $\mu^*$=0.13 (solid red line), our anharmonic $T_c$ of ~113 K at 190 GPa closely matches the highest experimental $T_c$ (113 K) reported at 191.7 GPa [7]. However, it is important to note that the same experiment also reported lower $T_c$ values of ~97 K at 171 GPa and ~107 K at 194 GPa. While our monotonic anharmonic curve does not capture the single off-trend data point at 194 GPa (~97 K), such a sharp nonmonotonic drop is atypical for a single-phase superconducting transition. As also noted in Ref. [7], this may arise from experimental pressure inhomogeneity, sample purity variations, or the possible presence of secondary phases (e.g., $YH_4$) in the low-pressure region. The excellent agreement at 190 GPa, where the most reliable high-pressure data are reported, strongly validates our anharmonic model. In contrast, the harmonic approximation at the same pressure of 190 GPa on the quantum structure yields a substantially higher $T_c$ exceeding 135 K (black solid line), clearly overestimating the experimental value. It is crucial to emphasize that this harmonic value is based on the quantum structure. As a consistency check, we performed additional harmonic calculations on the classical structure and obtained $T_c$ = 109.8 K at 140 GPa, which is in excellent agreement with the 110 K reported by Shi et al. [6] (blue diamonds). Therefore, the large discrepancy between our harmonic results (black lines) and the data of Shi et al. does not stem from numerical differences, but rather from the fact that our harmonic calculations are performed on a structurally renormalized configuration. This comparison clearly illustrates that the structure optimization pathway (harmonic vs. anharmonic) has a decisive impact on the predicted superconducting properties. Interestingly, the harmonic approximation based on the classical structure (Method A)—over the extended pressure window from 120 to 190 GPa, our calculated $T_c$ exhibits a mild increase with pressure. This trend is somewhat unexpected, as a decreasing or nearly flat $T_c$ is more commonly observed for most superconducting hydrides in such a broad pressure range. The excellent quantitative agreement confirms that the anharmonic treatment accurately describes the lattice dynamics and electron–phonon coupling in this system. Further high-pressure experiments on well–characterized single‑phase samples are highly desirable to clarify the intrinsic pressure evolution of $T_c$ in this system. Our anharmonic calculations, however, predict a smooth and monotonic decrease of $T_c$ with increasing pressure once the stability range is entered.

Further insight into the anharmonic renormalization of superconductivity is provided by the pressure evolution of the electron-phonon coupling constant λ and the logarithmic average frequency $\omega_{log}$, shown in Fig. 3(b). The harmonic λ (black squares) is systematically higher than the anharmonic one (red squares) across the entire pressure range, dropping from about 2.38 at 160 GPa to 1.8 at 190 GPa, while the anharmonic $\omega_{log}$ (red circles) is higher than its harmonic counterpart (black circles). This indicates that the specific hydrogen-related phonon modes that couple strongly to the electrons are hardened by anharmonicity, suppressing λ and consequently lowering $T_c$ relative to the harmonic prediction at a given pressure. For comparison, the data from Shi et al. [6] (blue squares and circles) at 140 GPa show a λ value very close to our anharmonic result on the quantum structure at the same pressure, but their $\omega_{log}$ is larger by about 150 $cm^{-1}$. We have verified that our harmonic calculations on the classical structure reproduce their λ and $T_c$ values exactly (for instance, at 140 GPa we obtained λ = 1.53 and $T_c$ = 109.8 K, in excellent agreement with their λ = 1.54 and $T_c$ = 110 K at 140 GPa). Therefore, the close agreement between Shi et al.'s classical-structure harmonic λ and our quantum-structure anharmonic λ at 140 GPa is coincidental, as it does not arise from a consistent renormalization of the phonon spectrum. Remarkably, the anharmonic calculations reveal a maximum $T_c$ of 123 K at 100 GPa (with $\mu$*=0.13), after which $T_c$ gradually decreases as pressure increases. This behavior is in stark contrast to the harmonic calculations on the classical structure (Method A), where $T_c$ rises monotonically and slowly with pressure.

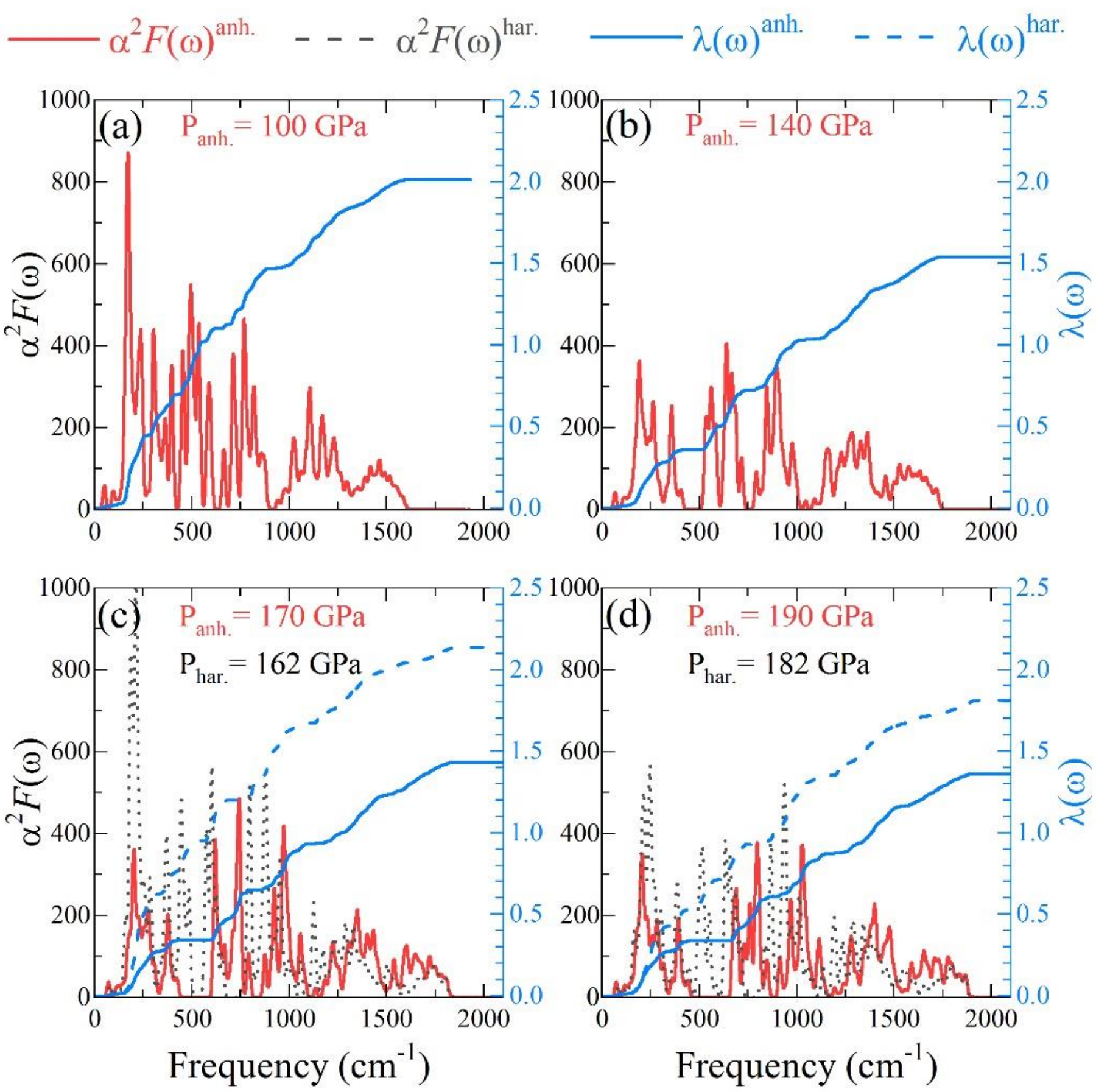

FIG. 4. Anharmonic spectral function $\alpha^2F(\omega)$ (red solid lines) and integrated electron-phonon coupling constant $\lambda(\omega)$ (blue solid lines) at four different anharmonic (quantum) pressures: (a) 100 GPa, (b) 140 GPa, (c) 170 GPa, and (d)

190 GPa, respectively. For comparison, harmonic results evaluated on the same quantum structure are also shown in panels (c) and (d) as dotted lines: black dotted for $\alpha^2F(\omega)$ and blue dotted for $\lambda(\omega)$, obtained at the corresponding harmonic pressures of 162 GPa (c) and 182 GPa (d), respectively. In panels (c) and (d), which contain both harmonic and anharmonic results, the two calculations are performed with the same lattice parameter.

Figure 4 quantifies the contributions to the total electron-phonon coupling constant λ from three frequency regions: low (<500 $cm^{-1}$), intermediate (500~1500 $cm^{-1}$), and high (>1500 $cm^{-1}$), obtained from both anharmonic (Anh) and harmonic (Har) calculations. At 190 GPa, the anharmonic λ values are 0.34 (low), 0.53 (mid), and 0.48 (high), summing to 1.35. The intermediate-frequency region, dominated by hydrogen bending modes, gives the largest contribution (~39%). As pressure increases from 100 to 190 GPa, the anharmonic total $\lambda$ decreases from 2.01 to 1.35, primarily due to a significant reduction of the low-frequency contribution (0.61 to 0.34) and the mid-frequency contribution (0.86 to 0.53), whereas the high-frequency part drops only modestly (0.54 to 0.48).

Crucially, a direct comparison between harmonic and anharmonic results at 170 and 190 GPa reveals a systematic overestimation of $\lambda$ by the harmonic approximation. At 190 GPa, the harmonic values are 0.51 (low), 0.90 (mid), and 0.40 (high), yielding a total λ of 1.81, which is 34 % higher than the anharmonic total of 1.35. The most pronounced discrepancy appears in the mid-frequency region, where the harmonic contribution (0.90) is nearly 70 % larger than its anharmonic counterpart (0.53). This trend is consistent with the phonon dispersion shown in Fig. 2, where harmonic frequencies are generally lower than their anharmonic counterparts. It is well known that phonon softening generally enhances the electron-phonon coupling and the superconducting critical temperature. Consequently, the harmonic total λ at 190 GPa (1.81) and 170 GPa (2.14) is substantially higher than the anharmonic values (1.35 and 1.42, respectively). This overestimation of λ inevitably leads to an overestimation of the harmonic superconducting critical temperature $T_c$. Thus, the harmonic approximation not only overestimates the electron-phonon coupling but also yields a substantially higher $T_c$ compared to the experimental value [7]. In contrast, the anharmonic λ values obtained here, combined with the logarithmic average frequencies derived from $\alpha^2F(\omega)$, yield $T_c$ in excellent agreement with the experimental measurements[7]. This anharmonic suppression of the total λ, particularly in the mid-frequency range, raises a crucial question: which specific atomic sites and vibrational modes are responsible for this renormalization? To address this, we decompose the Eliashberg spectral function and the phonon density of states into atomic contributions in Figs. 5 and 6, respectively.

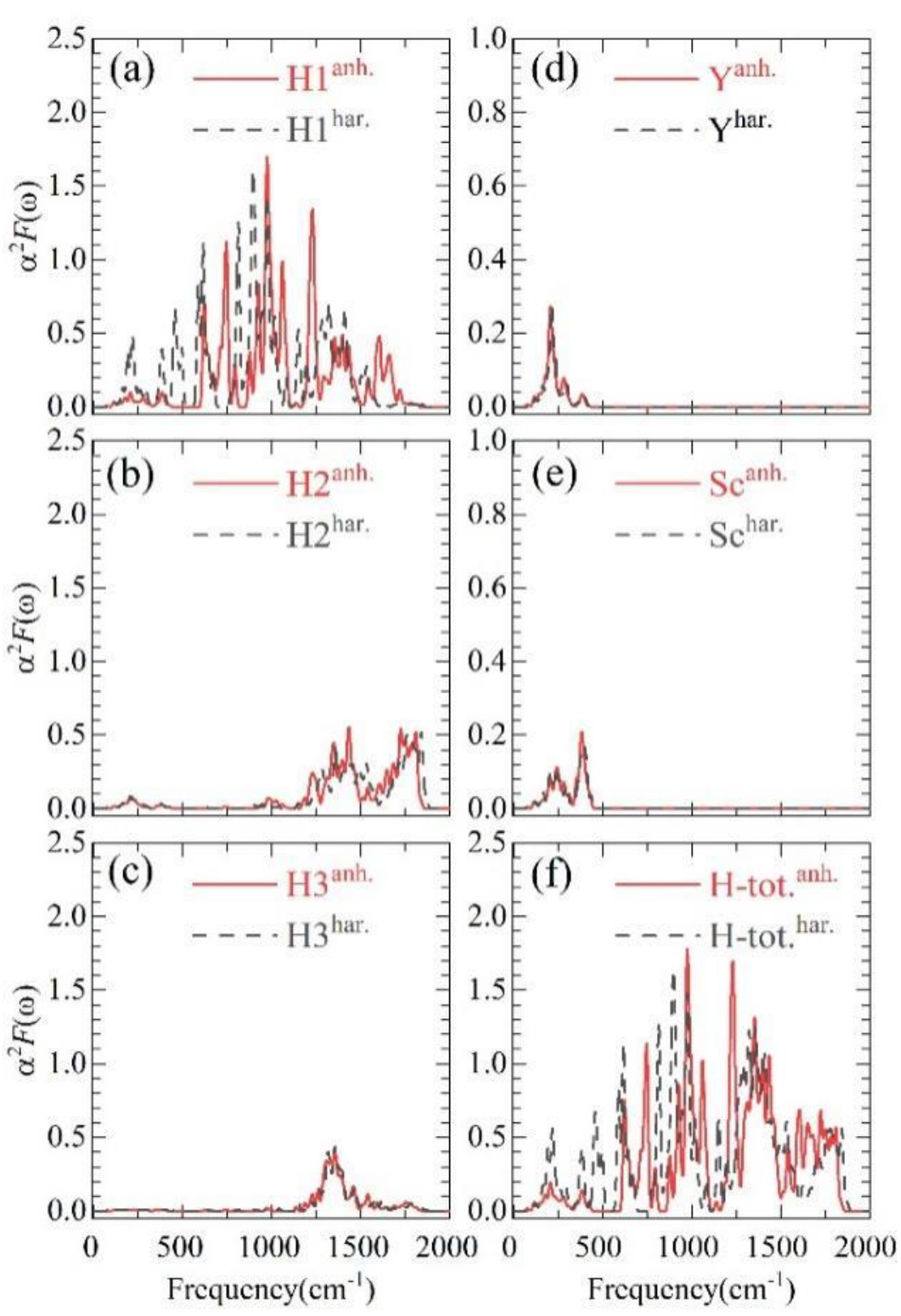

FIG. 5. Anharmonic (red solid lines) and harmonic evaluated on the quantum structure (black dotted lines) electron-phonon spectral functions α²$F$(ω) decomposed into contributions from different elements: (a) H1, (b) H2, (c) H3, (d) Y, (e) Sc. Panel (f) shows the total α²$F$(ω) from the sum of H1, H2, and H3.

The Eliashberg spectral function $\alpha^2F(\omega)$ decomposed into contributions from different elements at 170 GPa are displayed in Fig. 5. The contributions from each specific atom to $\alpha^2F(\omega)$, obtained by counting only the contribution from the atom itself in $\alpha^2F(\omega) = \sum_{\bar{a}\bar{b}} \alpha^2 F_{\bar{a}\bar{b}}(\omega)$ $[\alpha^2 F_{\bar{a}\bar{b}}(\omega)$ can be trivially obtained from Eqs. (13)]. Moreover, the phonon density of states (PDOS) and its atomic projections are also shown in Fig. 6. The contributions from each type of atoms to both $\alpha^2F(\omega)$ and PDOS are distinctly separated. From the figures, conclusions can be drawn. The difference between the anharmonic and harmonic total $\alpha^2F(\omega)$ spectra in Fig. 5(f) is overwhelmingly dominated by the contribution of the H1 atoms. This result is fully consistent with the preceding structural analysis: the anharmonicity–driven lattice distortion is predominantly confined to the edge H–H pairs ($d_1$) along the c–axis, and it is precisely the out–of–plane local vibrational modes of these H1 atoms that dominate the anharmonic correction to the electron–phonon coupling. In contrast, the contributions from H2, H3, and the metal atoms (Y and Sc) to the anharmonic response are negligible, indicating that in this superhydride, anharmonic effects do not act uniformly over the entire lattice, but rather modulate the electron–phonon interaction through the local vibrations of hydrogen atoms at specific interstitial sites. This finding provides a clear atomic–scale picture for understanding the influence of anharmonicity on the superconducting properties of compressed superhydrides: the anharmonic correction to the critical temperature primarily arises from those hydrogen sites that exhibit the most pronounced quantum structural distortions (such as the H1 site in the present system), while the harmonic approximation overestimates the electron–phonon coupling strength contributed by these localized vibrational modes, thus systematically overestimating $T_c$.

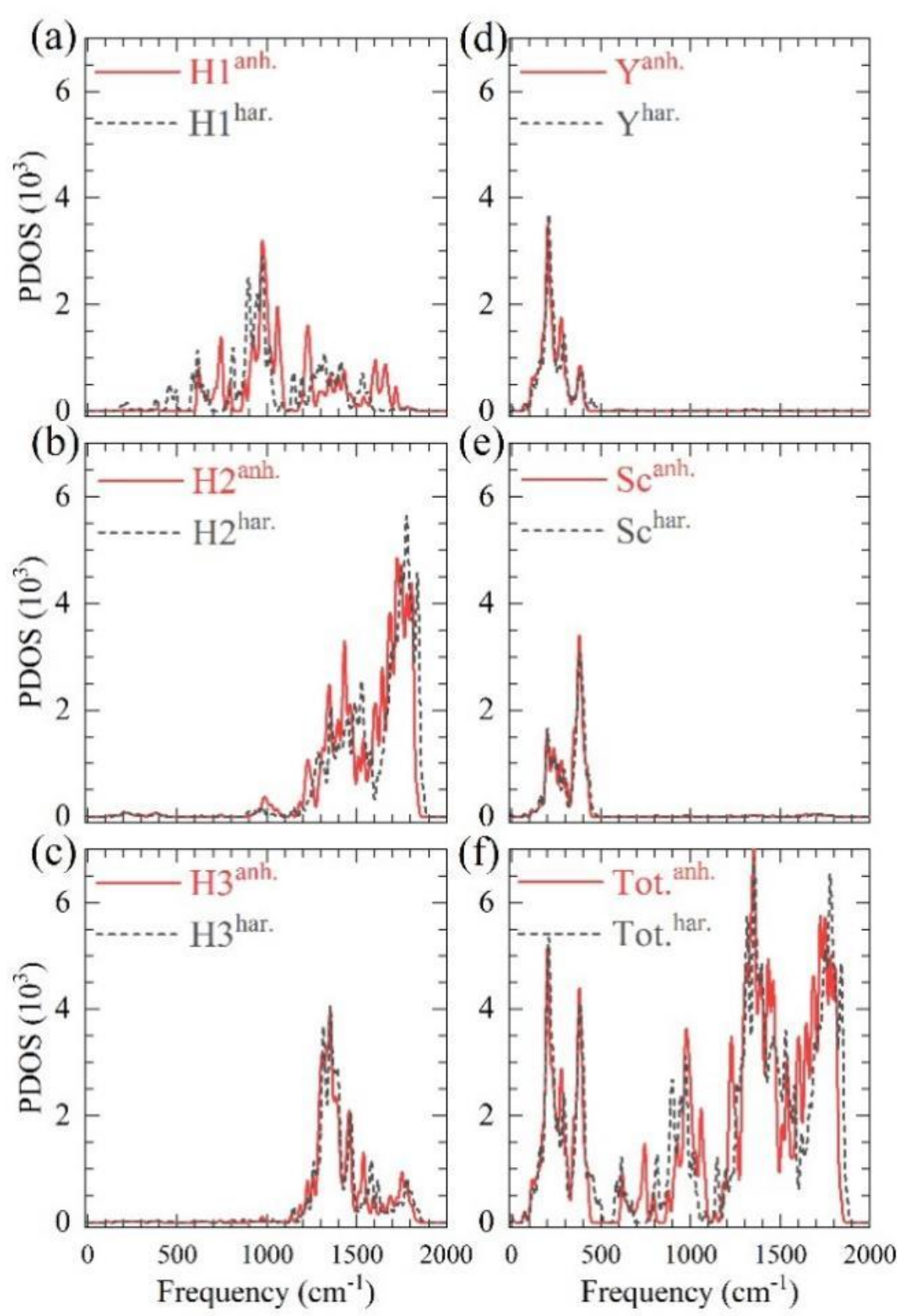

FIG. 6. Anharmonic (red solid lines) and harmonic evaluated on the quantum structure (black dotted lines) phonon density of states (PDOS) decomposed into contributions from different elements: (a) H1, (b) H2, (c) H3, (d) Y, (e) Sc. Panel (f) shows the total PDOS from the sum of H1, H2, and H3 contributions.

Figure 6 displays the atom‑projected phonon density of states (PDOS) for *P4/mmm*–$YScH_8$, comparing anharmonic (red solid) and harmonic (black dotted) calculations. As in the spectral–function decomposition of Fig. 5, the anharmonic effects on the PDOS are highly site-selective. For H1 [panel (a)], the anharmonic treatment induces marked modifications in both the peak positions and the spectral weight distribution over the entire frequency range, while the PDOS of H2 and H3 [panels (b) and (c)] remain nearly indistinguishable between the two approximations. Likewise, the contributions from Y and Sc [panels (d) and (e)] are essentially unchanged, consistent with the heavy masses of these metal atoms, which suppress their zero‑point vibrational amplitudes. As a result, the total PDOS [panel (f)] inherits its entire anharmonic deviation exclusively from the H1 atoms.

The PDOS directly reflects the vibrational density of states weighted by atomic displacements. Among all hydrogen sites, the H1–related phonon modes exhibit the most pronounced anharmonic reshaping, which is fully consistent with the structural analysis in Fig. 1, where the edge H–H distance $d_1$ shows the largest anharmonic distortion. This correspondence confirms that the local vibrational environment of the H1 site is the most sensitive to anharmonicity. Furthermore, the modified phonon spectrum of H1 is also in line with the changes observed in the Eliashberg spectral function $\alpha^2F(\omega)$, reinforcing that the anharmonic effects on lattice dynamics and electron–phonon coupling are primarily localized at the H1 site.

## V. CONCLUSIONS

In summary, we have systematically reinvestigated the structural stability and superconducting properties of the recently synthesized ternary superhydride *P4/mmm*-$YScH_8$ by explicitly including quantum ionic fluctuations and lattice anharmonicity via the stochastic self-consistent harmonic approximation. Our work demonstrates three key findings. First, including anharmonicity does not merely renormalize the phonon frequencies, but fundamentally alters the equilibrium crystal structure itself. We show that harmonic calculations performed on this anharmonically-renormalized "quantum structure" differ drastically from previous harmonic results obtained on the classical structure, and that only the anharmonic calculations on the quantum structure can reconcile theory with experiment. Second, the dynamical stability threshold is extended from ~145 GPa (harmonic) down to ~84 GPa (anharmonic), predicting a much wider pressure window for potential synthesis. Third, the anharmonic phonon spectrum yields a $T_c$ of ~113 K at 190 GPa, in excellent agreement with the highest experimental value, and reveals a monotonic decrease of $T_c$ with pressure above 100 GPa. More broadly, our work establishes that for ternary hydrides, site-selective anharmonicity **-** particularly at specific hydrogen interstitial sites **-** can govern the superconducting properties. This underscores that a consistent anharmonic treatment of both structure and lattice dynamics is indispensable for accurate predictions, and provides a refined theoretical framework for guiding future experimental and computational efforts in the search for high-temperature superconductors.

## ACKNOWLEDGMENTS

P.H. thankfully acknowledges the International Science and Technology Cooperation Project of Jilin Provincial Department of Science and Technology (Grant No. 20240402055GH).